# Improvement of the Han-Kobayashi Rate Region for General Interference Channel-v2

Ghosheh Abed Hodtani
Department of Electrical Engineering, Ferdowsi University of Mashhad
ghodtani@gmail.com

**Abstract.** Allowing the input auxiliary random variables to be correlated and using the binning scheme, the Han-Kobayashi (HK) rate region for general interference channel is partially improved. The obtained partially new achievable rate region (i) is compared to the HK region and its simplified description, i.e., Chong-Motani-Garg (CMG) region, in a detailed and favorable manner, by considering different versions of the regions, and (ii) has an interesting and easy interpretation: as expected, any rate in our region has generally two additional terms in comparison with the HK region (one due to the input correlation and the other as a result of the binning scheme).

## I. INTRODUCTION

Interference channel (IC) has been the most important and complicated channel for information theory researchers since its initiation by Shannon [1]; and is recently being studied in great detail due to its wide range of potential applications. Here we consider only the two user IC, where each sender communicates with its respective receiver interfering with communication of the other sender-receiver.

The study of the IC was furthered by Ahlswede [2]. Sato [3] obtained various inner and outer bounds by considering the associated multiple access sub-channel in the IC. Carleial [4] established an improved achievable rate region (with one auxiliary random variable for each sender ) by using sequential decoding and convex hull operation based on the superposition coding of Cover [5].

Han and Kobayashi (KH) [6],[7], generalized Cover's superposition coding to the many variable case; applied jointly or simultaneous decoding strategy instead of sequential decoding in [4] and the time-sharing formulation instead of convex hull operation in [4] for the general IC, thereby establishing the most popular achievable strategy and the best achievable rate region known to date. Chong, Motani and Garg [8], by slightly modifying the decoding error definition and reducing the number of auxiliary random variables by superposition coding, derived a simplified description for the HK rate region, which we will refer to as the CMG region. In [8],[7], the equivalence of the regions is proved.

With the exception of a few special cases, the capacity region of the IC is not known. The problem of determining the capacity region and even some rate regions has been studied dominantly from the viewpoint of previously investigated special cases of multiple and broadcast sub-channels in the IC: [6],[9], and [10]-[20].

The Gaussian IC has been intensively studied in [6],[10],[21]-[27].

In this paper, first, we improve the HK region, by allowing the input auxiliary random variables at each transmitter to be correlated and using the Gelfand-Pinsker binning scheme [28] as in Marton coding for broadcast channel [29], and the HK celebrated jointly decoding strategy [6]. Then, we show that the HK region and hence its simplified description, i.e., the CMG region are special cases of our region, by providing a detailed comparison between different versions of the regions.

The remainder of the paper is as follows. In section II, we define the IC and the modified IC. In section III, the HK and the CMG regions are recalled and their different versions are derived, explained and compared to each other. Then, in section IV, we obtain our new region for the IC, derive its different descriptions and compare it to the HK and the CMG regions. Finally, a conclusion is prepared in section V.

## II. DEFINITIONS

We denote random variables by $X_1, X_2, Y_1, \cdots$ with values $x_1, x_2, y_1, \cdots$ in finite sets $\mathcal{X}_1, \mathcal{X}_2, \mathcal{Y}_1, \cdots$ respectively; n-tuple vectors of $X_1, X_2, Y_1, \cdots$ are denoted with $\boldsymbol{x_1, x_2, y_1}, \cdots$. We use the symbol $A_\varepsilon^n(X_1, X_2, \cdots, X_l)$ to indicate the set of $\varepsilon$-typical n-sequences $(\boldsymbol{x_1, x_2, \cdots, x_l})$ [30].

**Interference Channel (IC)**

A discrete and memoryless IC $(\mathcal{X}_1 \times \mathcal{X}_2, p(y_1 y_2 | x_1 x_2), \mathcal{Y}_1 \times \mathcal{Y}_2)$ consists of two sender-receiver pairs $(X_1 \rightarrow Y_1$ and $X_2 \rightarrow Y_2)$ in Fig.1, where $\mathcal{X}_1, \mathcal{X}_2$ are two finite input alphabet sets; $\mathcal{Y}_1, \mathcal{Y}_2$ are two finite output alphabet sets, and $p(y_1 y_2 | x_1 x_2)$ is a conditional channel probability of $(y_1, y_2) \in \mathcal{Y}_1 \times \mathcal{Y}_2$ given $(x_1, x_2) \in \mathcal{X}_1 \times \mathcal{X}_2$. The nth extension of the channel is:

$$p(\boldsymbol{y_1 y_2 | x_1 x_2}) = \prod_{i=1}^{n} p(y_{1i} y_{2i} | x_{1i} x_{2i})$$

---

A code $(n, M_1 = \lfloor 2^{nR_1} \rfloor, M_2 = \lfloor 2^{nR_2} \rfloor, \varepsilon)$ is a collection of $M_1$ codewords $\boldsymbol{x_{1i}} \in \mathcal{X}_1^n, i \in \mathcal{M}_1$; $M_2$ codewords $\boldsymbol{x_{2j}} \in \mathcal{X}_2^n, j \in \mathcal{M}_2$; two decoding functions $g_1: \boldsymbol{y_1} \rightarrow \mathcal{M}_1, g_2: \boldsymbol{y_2} \rightarrow \mathcal{M}_2$; and the average error probabilities at the receivers $(P_{e_1}^n, P_{e_2}^n)$ are defined conveniently [6],[8].

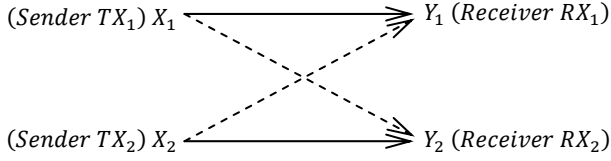 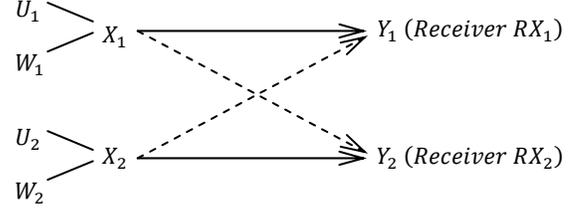

Fig.1 Interference channel      Fig.2 Modified interference channel

A pair $(R_1, R_2)$ of non-negative real values is called an achievable rate if there exists a sequence of codes such that under some decoding scheme, $\max(P_{e_1}^n, P_{e_2}^n) < \varepsilon$.

The capacity region of the IC is the set of all achievable rates.

**Modified interference channel**

As in [6], a modified IC (Fig.2), models two senders communicating both private and common message to two receivers; where the information conveying role of the channel inputs $X_1, X_2$ is transferred to some fictitious inputs $U_1, W_1, U_2, W_2$, so that the channel behaves like a channel $U_1, W_1, U_2, W_2 \rightarrow Y_1 Y_2$.

Auxiliary random variables $W_1$ and $W_2$ represent the public message to be sent from $TX_1$ to $(RX_1, RX_2)$ with the rate $T_1$ and from $TX_2$ to $(RX_1, RX_2)$ with the rate $T_2$, respectively. Similarly, $U_1$ and $U_2$ are the private message to be sent from $TX_1$ to $RX_1$ with the rate $S_1$ and from $TX_2$ to $RX_2$ with the rate $S_2$, respectively. Also, as in [6], $Q \in \mathcal{Q}$ is a time sharing random variable whose n-sequences $\boldsymbol{q} = (q_1, q_2, \cdots, q_n)$ are generated independently of the messages. The n-sequences $\boldsymbol{q}$ are given to both senders and receivers.

An $(n, \lfloor 2^{nT_1} \rfloor, \lfloor 2^{nS_1} \rfloor, \lfloor 2^{nT_2} \rfloor, \lfloor 2^{nS_2} \rfloor, \varepsilon)$ code for the modified IC (Fig.2) consists of $\lfloor 2^{nT_1} \rfloor$ codewords $\boldsymbol{w_1}(j)$, $\lfloor 2^{nS_1} \rfloor$ codewords $\boldsymbol{u_1}(l)$ for $TX_1$; and $\lfloor 2^{nT_2} \rfloor$ codewords $\boldsymbol{w_2}(m)$, $\lfloor 2^{nS_2} \rfloor$ codewords $\boldsymbol{u_2}(k)$ for $TX_2$; $j \in \{1, \cdots, 2^{nT_1}\}$, $l \in \{1, \cdots, 2^{nS_1}\}$, $m \in \{1, \cdots, 2^{nT_2}\}$, $k \in \{1, \cdots, 2^{nS_2}\}$, such that the maximum of the conveniently defined average probabilities of decoding error $(P_{e_1}^n, P_{e_2}^n)$ is less than $\varepsilon$.

A quadruple $(T_1, S_1, T_2, S_2)$ of non-negative real numbers is achievable for the modified IC (and hence, $(R_1 = S_1 + T_1, R_2 = S_2 + T_2)$ is achievable rate for the IC) if there exists a sequence of codes such that the maximum of average error probabilities under some decoding scheme is less than $\varepsilon$. An achievable region for the modified IC is the closure of a subset of the positive region $R^4$ of achievable rate quadruples $(T_1, S_1, T_2, S_2)$.

Therefore, we can consider auxiliary random variables $Q, U_1, W_1, U_2, W_2$, defined on arbitrary finite sets $\mathcal{Q}, \mathcal{U}_1, \mathcal{W}_1, \mathcal{U}_2, \mathcal{W}_2$, respectively; $X_1$ and $X_2$ defined on the input alphabet sets $\mathcal{X}_1, \mathcal{X}_2$, and $Y_1, Y_2$, defined on the output alphabet sets $\mathcal{Y}_1$ and $\mathcal{Y}_2$. Let $Z = (QU_1 W_1 U_2 W_2 X_1 X_2 Y_1 Y_2)$ and let $\mathcal{P}_{IC}$ be the set of all distributions of the form (for Fig.2): (hereafter, for brevity, let $p(qu_1 w_1 u_2 w_2 x_1 x_2 y_1 y_2) = p(\ )$)

$$p(\ ) = \boldsymbol{p(q)p(u_1 w_1|q)p(u_2 w_2|q)}p(x_1|qu_1 w_1)p(x_2|qu_2 w_2)p(y_1 y_2|x_1 x_2) \qquad (1).$$

### III. THE HK AND THE CMG REGIONS

**A. The HK rate region**

Han and Kobayashi [6] considered the general distribution (1) in a special case of the form:

$$p(\ ) = \boldsymbol{p(q)p(w_1|q)p(u_1|q)p(w_2|q)p(u_2|q)}p_q(y_1 y_2|u_1 w_1 u_2 w_2) \qquad (2),$$

where $p_q(y_1 y_2|u_1 w_1 u_2 w_2) = p(y_1 y_2|x_1 = f_1(u_1 w_1|q), x_2 = f_2(u_2 w_2|q))$

and by using superposition coding of $w_1, u_1, w_2, u_2$ over $q$ and jointly decoding strategy, derived the best achievable rate region known to date as follows. In the HK coding, the common and private message codewords $(\boldsymbol{q}, \boldsymbol{u_1}, \boldsymbol{w_1}), (\boldsymbol{q}, \boldsymbol{u_2}, \boldsymbol{w_2})$ are mapped into signal spaces $(\boldsymbol{x_1}), (\boldsymbol{x_2})$, through arbitrary deterministic functions $f_1, f_2$ respectively, [6].

**Theorem 1** ([6], theorem 3.1): For the modified IC (Fig.2), let $Z = (QU_1 W_1 U_2 W_2 X_1 X_2 Y_1 Y_2)$ and let $\mathcal{P}_{IC}^{HK}$ be the set of all distributions of the special form (2). For any $Z \in \mathcal{P}_{IC}^{HK}$ let $S_{IC}^{HK}(Z)$ be the set of all quadruples $(T_1, S_1, T_2, S_2)$ of non-negative real numbers such that

$$S_1 \leq I(Y_1; U_1|W_1 W_2 Q) = a_1 \qquad (3\text{-}1),$$

$$T_1 \leq I(Y_1; W_1|U_1 W_2 Q) = b_1 \qquad (3\text{-}2),$$

$$T_2 \leq I(Y_1; W_2|U_1 W_1 Q) = c_1 \qquad (3\text{-}3),$$

$$S_1 + T_1 \leq I(Y_1; U_1 W_1|W_2 Q) = d_1 \qquad (3\text{-}4),$$

$$S_1 + T_2 \leq I(Y_1; U_1 W_2|W_1 Q) = e_1 \qquad (3\text{-}5),$$

$$T_1 + T_2 \leq I(Y_1; W_1 W_2 | U_1 Q) = f_1 \quad (3\text{-}6),$$
$$S_1 + T_1 + T_2 \leq I(Y_1; U_1 W_1 W_2 | Q) = g_1 \quad (3\text{-}7),$$
$$S_2 \leq I(Y_2; U_2 | W_1 W_2 Q) = a_2 \quad (3\text{-}8),$$
$$T_2 \leq I(Y_2; W_2 | U_2 W_1 Q) = b_2 \quad (3\text{-}9),$$
$$T_1 \leq I(Y_2; W_1 | U_2 W_2 Q) = c_2 \quad (3\text{-}10),$$
$$S_2 + T_2 \leq I(Y_2; U_2 W_2 | W_1 Q) = d_2 \quad (3\text{-}11),$$
$$S_2 + T_1 \leq I(Y_2; U_2 W_1 | W_2 Q) = e_2 \quad (3\text{-}12),$$
$$T_1 + T_2 \leq I(Y_2; W_1 W_2 | U_2 Q) = f_2 \quad (3\text{-}13),$$
$$S_2 + T_1 + T_2 \leq I(Y_2; U_2 W_1 W_2 | Q) = g_2 \quad (3\text{-}14),$$

then any element of the closure of $\bigcup_{Z \in \mathcal{P}_{IC}^{HK}} S_{IC}^{HK}(Z)$ is achievable.

**Proof.** Refer to [6].

**Note.** Hereafter, $a_i, b_i, c_i, d_i, e_i, f_i$ and $g_i$, $i = 1, 2$ are the same as in theorem 1, unless otherwise stated.

**Amended version of theorem B in [7]**

Now, we transform the above region into the rate pair ($R_1 = S_1 + T_1, R_2 = S_2 + T_2$) using the Fourier-Motzkin elimination technique and apply the independence of $U_i$ and $W_i$ given $Q$, $i = 1, 2$ in the distribution (2) to the results and amend theorem B in [7].

**Theorem 2** [amended version of theorem B in [7]]: The region in theorem 1 can be described as $\mathcal{R}_{HK}$ being the set of ($R_1$, $R_2$) satisfying:

$$R_1 \leq d_1 \quad (4\text{-}1),$$
$$\mathbf{R_1 \leq a_1 + c_2} \quad (4\text{-}2),$$
$$R_2 \leq d_2 \quad (4\text{-}3),$$
$$\mathbf{R_2 \leq a_2 + c_1} \quad (4\text{-}4),$$
$$R_1 + R_2 \leq a_1 + g_2 \quad (4\text{-}5),$$
$$R_1 + R_2 \leq a_2 + g_1 \quad (4\text{-}6),$$
$$R_1 + R_2 \leq e_1 + e_2 \quad (4\text{-}7),$$
$$2R_1 + R_2 \leq a_1 + g_1 + e_2 \quad (4\text{-}8),$$
$$2R_2 + R_1 \leq a_2 + g_2 + e_1 \quad (4\text{-}9),$$

where $a_i, b_i, c_i, d_i, e_i, f_i$ and $g_i$, $i = 1, 2$ are the same as in theorem 1.

**Proof.** Refer to the proof of theorem B in [7]. However, in theorem B [7] there are two additional inequalities:
$$2R_1 + R_2 \leq 2a_1 + e_2 + f_2 \quad (4\text{-}10)$$
$$2R_2 + R_1 \leq 2a_2 + e_1 + f_1 \quad (4\text{-}11).$$

The inequalities (4-10,11) are obtained from (4-2,5) and (4-4,6), respectively, as follows and hence are redundant, merely as a result of the independence of $U_i$ and $W_i$ given $Q, i = 1,2$ in (2):

$$(4\text{-}2) + (4\text{-}5) \Rightarrow 2R_1 + R_2 \leq 2a_1 + c_2 + g_2 \quad (5),$$

and the independence of $U_2$ and $W_2$ given $Q$ results in $I(Y_2; U_2|Q) \leq I(Y_2; U_2|QW_2)$ (6), from where we have:
$$c_2 + g_2 = I(Y_2; W_1|W_2 U_2 Q) + I(Y_2; U_2 W_1 W_2 | Q) \leq e_2 + f_2 = I(Y_2; U_2 W_1 | W_2 Q) + I(Y_2; W_1 W_2 | Q U_2) \quad (7).$$

Therefore, in accordance with (7), the relation (5) yields (4-10), i.e., (4-10) is redundant. Similarly, (4-4,6) results in the redundancy of (4-11).

**Remark 1.** The fact that we have considered in theorem 2 is the intrinsic independence of $U_i$ and $W_i$ given $Q$, $i = 1,2$ in the HK region.

**Lemma 1.** Explanatory and easily comparable form of $\mathcal{R}_{HK}$ in theorem 2 can be described as ($R_1$, $R_2$) satisfying thirteen relations consisting of (4-1,...11) and $R_1 \leq a_1 + e_2$ (4-12) and $R_2 \leq a_2 + e_1$ (4-13).

**Proof.** In view of $c_i \leq e_i$, $i = 1, 2$, (4-2) and (4-4) yields (4-12) and (4-13), respectively. Therefore, ($R_1$, $R_2$) satisfies (4-1,...13).

**Theorem 3** (theorem C in [7]). Assuming that the incorrect decoding of $W_1(W_2)$ by the receiver $RX2(RX1)$ is not considered as an error, the region in theorem1 for the modified IC in Fig.2, is changed as follows.

$$R_1 \leq d_1 \quad (8\text{-}1)$$
$$R_1 \leq a_1 + e_2 \quad (8\text{-}2)$$
$$R_1 \leq a_1 + f_2 \quad (8\text{-}3)$$
$$R_2 \leq d_2 \quad (8\text{-}4)$$
$$R_2 \leq a_2 + e_1 \quad (8\text{-}5)$$

$$R_2 \leq a_2 + f_1 \tag{8-6}$$
$$R_1 + R_2 \leq a_2 + g_1 \tag{8-7}$$
$$R_1 + R_2 \leq a_1 + g_2 \tag{8-8}$$
$$R_1 + R_2 \leq e_1 + e_2 \tag{8-9}$$
$$2R_1 + R_2 \leq a_1 + g_1 + e_2 \tag{8-10}$$
$$2R_1 + R_2 \leq 2a_1 + e_2 + f_2 \tag{8-11}$$
$$2R_2 + R_1 \leq a_2 + g_2 + e_1 \tag{8-12}$$
$$2R_2 + R_1 \leq 2a_2 + e_1 + f_1 \tag{8-13}$$

**Proof.** We apply the Fourier-Motzkin algorithm to the region in theorem 1 without the inequalities (3-3,10) in the same manner as in theorem 2. For brevity the details are omitted.

**B. The CMG rate region**

Chong, Motani and Garg [8]:

1. Modified the error definition slightly, hence considered the relations (3-3,10) unnecessary in evaluating the error probabilities.

2. Reduced the number of auxiliary random variables and used $Q, W_1, W_2$ instead of $Q, U_1, W_1, U_2, W_2$, while superimposing the message conveyed by $U_1 (U_2)$ over $Q, W_1(Q, W_2)$ by $X_1(X_2)$. In other words, they considered a kind of correlation between the inputs, by removing $(u_1, u_2)$ and substituting them by $(x_1, x_2)$, through the following distribution:

$$p() = p(q)\, p(w_1|q)\, p(x_1|qw_1)\, p(w_2|q)\, p(x_2|qw_2) p(y_1y_2|x_1x_2) \tag{9}$$

3. Using modified error definition, superposition coding in accordance with (9), the HK jointly decoding strategy and considering the Markov chains [ $W_1 \to QW_2X_1 \to Y_1$ (10), $W_2 \to QW_1X_2 \to Y_2$ (11) ] resulting from (9), they removed the inequalities related to the rates $T_1, T_2$ and $T_1 + T_2$ in the region of theorem 1 and derived the following region as follows.

**Theorem 4** ( lemma 3, [8] ). For the modified IC in Fig. 2, let $Z_1 = (QW_1W_2X_1X_2Y_1Y_2)$ and let $\mathcal{P}_{IC}^{CMG}$ be the set of all distributions of the form (9). For any $Z_1 \in \mathcal{P}_{IC}^{CMG}$ let $S_{IC}^{CMG}(Z_1)$ be the set of all quadruples $(T_1, S_1, T_2, S_2)$ of non-negative real numbers such that

$$S_1 \leq I(Y_1; X_1|W_1W_2Q) = \mathcal{A}_1 \tag{12-1}$$
$$S_1 + T_1 \leq I(Y_1; X_1|W_2Q) = \mathcal{D}_1 \tag{12-2}$$
$$S_1 + T_2 \leq I(Y_1; X_1W_2|W_1Q) = \mathcal{E}_1 \tag{12-3}$$
$$S_1 + T_1 + T_2 \leq I(Y_1; X_1W_2|Q) = \mathcal{G}_1 \tag{12-4}$$
$$S_2 \leq I(Y_2; X_2|W_2W_1Q) = \mathcal{A}_2 \tag{12-5}$$
$$S_2 + T_2 \leq I(Y_2; X_2|W_1Q) = \mathcal{D}_2 \tag{12-6}$$
$$S_2 + T_1 \leq I(Y_2; X_2W_1|W_2Q) = \mathcal{E}_2 \tag{12-7}$$
$$S_2 + T_2 + T_1 \leq I(Y_2; X_2W_1|Q) = \mathcal{G}_2 \tag{12-8}$$

then, any element of the closure of $\bigcup_{Z_1 \in \mathcal{P}_{IC}^{CMG}} S_{IC}^{CMG}(Z_1)$ is achievable.

**Proof.** Refer to [8]. To explain the proof in [8], it is worth noting that the relations (3-3,10) in theorem 1 are removed due to modifying the error definition. Also, besides (12-1,…,8), we have four relations in the proof [8]:

$$T_i \leq \mathcal{D}_i, \quad i = 1,2 \tag{12-9,10}$$
$$T_1 + T_2 \leq \mathcal{G}_i, i = 1,2, \tag{12-11,12}$$

which become redundant due to (12-2,6) and (12-4,8) respectively. Therefore, for the CMG coding, 14 inequalities in the HK region are reduced to 8 inequalities (12-1,…8).

**Lemma 2.** For $a_i, d_i, e_i, g_i$, $i = 1,2$ in theorem 1 and $\mathcal{A}_i, \mathcal{D}_i, \mathcal{E}_i, \mathcal{G}_i$, $i = 1,2$ in theorem 4, we have generally:

$$a_1 = I(Y_1; U_1|W_1W_2Q) \geq I(Y_1; X_1|W_1W_2Q) = \mathcal{A}_1 \tag{13-1}$$
$$d_1 = I(Y_1; U_1W_1|W_2Q) \geq I(Y_1; X_1|W_2Q) = \mathcal{D}_1 \tag{13-2}$$
$$e_1 = I(Y_1; U_1W_2|W_1Q) \geq I(Y_1; X_1W_2|W_1Q) = \mathcal{E}_1 \tag{13-3}$$
$$g_1 = I(Y_1; U_1W_1W_2|Q) \geq I(Y_1; X_1W_2|Q) = \mathcal{G}_1 \tag{13-4}$$
$$a_2 = I(Y_2; U_2|W_2W_1Q) \geq I(Y_2; X_2|W_2W_1Q) = \mathcal{A}_2 \tag{13-5}$$
$$d_2 = I(Y_2; U_2W_2|W_1Q) \geq I(Y_2; X_2|W_1Q) = \mathcal{D}_2 \tag{13-6}$$
$$e_2 = I(Y_2; U_2W_1|W_2Q) \geq I(Y_2; X_2W_1|W_2Q) = \mathcal{E}_2 \tag{13-7}$$
$$g_2 = I(Y_2; U_2W_2W_1|Q) \geq I(Y_2; X_2W_1|Q) = \mathcal{G}_2 \tag{13-8}$$

that, equalities (=) hold for the corresponding distributions (2) and (9).

**Proof.** Due to the inequality $(H(Y|X) = H(Y|X, f(X)) \leq H(Y|f(X)))$, for the HK coding with encoding functions $X_1 = f_1(U_1, W_1|Q)$, we have: $H(Y_1|W_1W_2QX_1) \geq H(Y_1|W_1W_2QU_1W_1)$, and then, $a_1 = I(Y_1; U_1|W_1W_2Q) = H(Y_1|W_1W_2Q) - H(Y_1|W_1W_2QU_1) \geq H(Y_1|W_1W_2Q) - H(Y_1|W_1W_2QX_1) = I(Y_1; X_1|W_1W_2Q) = \mathcal{A}_1$; and similarly $a_2 \geq \mathcal{A}_2$ and (13-2)-(13-8) are proved.

As explained in [8] (p. 3190, the relation 29), due to the independence of $U_i$ from $W_i$, $i = 1,2$ in the HK region, there are always corresponding distributions (2), (9) with results (10) and (11), resulting in the (=) in (13-1) - (13-8). Specifically, $X_1 = f_1(U_1, W_1|Q)$ and (10) yields $a_1 = \mathcal{A}_1$ and etc.

**Theorem 5** ( lemma 4,[8] and theorem D,[7] ). By using the Fourier-Motzkin algorithm, the CMG region in theorem 4 can be described as $\mathcal{R}_{CMG}$ being the set of $(R_1, R_2)$ satisfying:

$$R_1 \leq \mathcal{D}_1 \tag{14-1}$$
$$\boldsymbol{R_1 \leq \mathcal{A}_1 + \mathcal{E}_2} \tag{14-2}$$
$$R_2 \leq \mathcal{D}_2 \tag{14-3}$$
$$\boldsymbol{R_2 \leq \mathcal{A}_2 + \mathcal{E}_1} \tag{14-4}$$
$$R_1 + R_2 \leq \mathcal{A}_1 + \mathcal{G}_2 \tag{14-5}$$
$$R_1 + R_2 \leq \mathcal{A}_2 + \mathcal{G}_1 \tag{14-6}$$
$$R_1 + R_2 \leq \mathcal{E}_1 + \mathcal{E}_2 \tag{14-7}$$
$$2R_1 + R_2 \leq \mathcal{A}_1 + \mathcal{G}_1 + \mathcal{E}_2 \tag{14-8}$$
$$2R_2 + R_1 \leq \mathcal{A}_2 + \mathcal{G}_2 + \mathcal{E}_1 \tag{14-9},$$

where $\mathcal{A}_i$, $\mathcal{D}_i$, $\mathcal{E}_i$, and $\mathcal{G}_i$, $i = 1, 2$ are the same as in theorem 4.

**Proof.** Refer to [7].

### C. Comparing the HK and the CMG regions

By comparing the two regions in view of $(T_1, S_1, T_2, S_2)$, it is seen that in theorem 1, the rates including $S_1$ and or $S_2$, (3-2,9), (3-6,13) give (12-1,…,8); (12-9,10), (12-11,12) in theorem 4, respectively. But the case is not true for the rates $T_1, T_2$ and $T_1 + T_2$.

The two regions should be compared in terms of $(R_1, R_2)$. It can be proved that the HK and the CMG $(R_1, R_2)$ regions are equivalent.

**Theorem 6..** The HK and the CMG $(R_1, R_2)$ regions are equivalent such that:

$(R_1, R_2) \in \mathcal{R}_{HK} = $ (4-1,…,9) or (4-1,…,13) $\Leftrightarrow (R_1, R_2) \in \mathcal{R}_{CMG} = $ (14-1,…,9) $\Leftrightarrow (R_1, R_2) \in $ (15-1,…,7),

where (15-1,…,7) are as follows.

$$R_1 \leq \mathcal{D}_1 \tag{15-1}$$
$$R_2 \leq \mathcal{D}_2 \tag{15-2}$$
$$R_1 + R_2 \leq \mathcal{A}_1 + \mathcal{G}_2 \tag{15-3}$$
$$R_1 + R_2 \leq \mathcal{A}_2 + \mathcal{G}_1 \tag{15-4}$$
$$R_1 + R_2 \leq \mathcal{E}_1 + \mathcal{E}_2 \tag{15-5}$$
$$2R_1 + R_2 \leq \mathcal{A}_1 + \mathcal{G}_1 + \mathcal{E}_2 \tag{15-6}$$
$$2R_2 + R_1 \leq \mathcal{A}_2 + \mathcal{G}_2 + \mathcal{E}_1 \tag{15-7}.$$

**Proof.** Due to the independence of $U_i$ from $W_i$, $i = 1,2$ in the HK region, there are always corresponding distributions (2) (with $X_i = f_i(U_i, W_i|Q)$, $i = 1,2$) and (9) (with results (10), (11)), thereby holding (=) in lemma 2 and resulting in the above equivalences, as proved in [8] ( theorem 2 for (15-1)-(15-7) ); lemma 1 for (4-1)(4-9); lemma 4 for (14-1)-(14-9); lemma 2 for proving the equivalency of the two regions).

**Remark 2.** The equivalency of the two regions is seen intuitively:

a. Not satisfying of both 4-2,4 (14-2,4) is contrary to 4-7 (14-7). And, when either 4-2 (14-2) or 4-4 (14-4) is not satisfied, it does not result in any contradiction to other inequalities.

b. In the CMG region, we don't have the terms $f_i$, $i = 1,2$. Therefore, removing the inequalities including $f_i$, $i = 1,2$ in the modified HK region, (8-1)-(8-13) leads to the CMG region (14-1)-(14-9).

## IV. NEW RATE REGION FOR GENERAL INTERFERENCE CHANNEL

Multiple access channels have been studied for independent [31],[32], specially correlated [33] and arbitrarily correlated [34] inputs. Taking into account these considerations, we study the IC in Fig.2 with a general input distribution and obtain a new achievable rate region by using binning scheme [28] and the HK jointly decoding strategy [6].

We, first, describe our region as the quadruple $(T_1, S_1, T_2, S_2)$ and then as $(R_1, R_2)$ rates. Finally, we compare our region with the HK and the CMG regions.

**Hodtani rate region for the IC**

Now, considering the general distribution (1) for the IC or allowing the auxiliary variables in the HK distribution (2) to be correlated in the following form:

$$\boldsymbol{p(\ ) = p(q)\, p(w_1|q)\, p(u_1|qw_1)\, p(w_2|q)\, p(u_2|qw_2) p_q(y_1 y_2|u_1 w_1 u_2 w_2)} \tag{16},$$

where $p_q(y_1 y_2|u_1 w_1 u_2 w_2) = p(y_1 y_2|x_1 = f_1(u_1 w_1|q), x_2 = f_2(u_2 w_2|q))$, we obtain a **partially improved** rate region as follows.

**Theorem 7.** For the modified IC (Fig.2), let $Z = (QU_1 W_1 U_2 W_2 X_1 X_2 Y_1 Y_2)$ and let $\mathcal{P}_{IC}^{Hod}$ be the set of all distributions of the form (16). For any $Z \in \mathcal{P}_{IC}^{Hod}$ let $S_{IC}^{Hod}(Z)$ be the set of all quadruples $(T_1, S_1, T_2, S_2)$ such that

$$S_1 \leq I(Y_1;U_1|W_1W_2Q) = a_1 \quad (17\text{-}1),$$
$$T_1 \leq I(Y_1;W_1|U_1W_2Q) + I(U_1;W_1|Q) = B_1 \quad (17\text{-}2),$$
$$T_2 \leq I(Y_1;W_2|U_1W_1Q) + I(U_1;W_1|Q) = C_1 \quad (17\text{-}3),$$
$$S_1 + T_1 \leq I(Y_1;U_1W_1|W_2Q) = d_1 \quad (17\text{-}4),$$
$$S_1 + T_2 \leq I(Y_1;U_1W_2|W_1Q) = e_1 \quad (17\text{-}5),$$
$$T_1 + T_2 \leq I(Y_1;W_1W_2|U_1Q) + I(U_1;W_1|Q) = F_1 \quad (17\text{-}6),$$
$$S_1 + T_1 + T_2 \leq I(Y_1;U_1W_1W_2|Q) = g_1 \quad (17\text{-}7),$$
$$S_2 \leq I(Y_2;U_2|W_1W_2Q) = a_2 \quad (17\text{-}8),$$
$$T_2 \leq I(Y_2;W_2|U_2W_1Q) + I(U_2;W_2|Q) = B_2 \quad (17\text{-}9),$$
$$T_1 \leq I(Y_2;W_1|U_2W_2Q) + I(U_2;W_2|Q) = C_2 \quad (17\text{-}10),$$
$$S_2 + T_2 \leq I(Y_2;U_2W_2|W_1Q) = d_2 \quad (17\text{-}11),$$
$$S_2 + T_1 \leq I(Y_2;U_2W_1|W_2Q) = e_2 \quad (17\text{-}12),$$
$$T_1 + T_2 \leq I(Y_2;W_1W_2|U_2Q) + I(U_2;W_2|Q) = F_2 \quad (17\text{-}13),$$
$$S_2 + T_1 + T_2 \leq I(Y_2;U_2W_1W_2|Q) = g_2 \quad (17\text{-}14),$$

then, any element of the closure of $\bigcup_{Z \in \mathcal{P}_{IC}^{Hod}} S_{IC}^{Hod}(Z)$ is achievable, where as seen, $a_i, d_i, e_i, g_i, i = 1,2$ are the same as in theorem 1 and $B_i, C_i, F_i, i = 1,2$ are new terms which are not equal to the corresponding terms $b_i, c_i, f_i, i = 1,2$ in theorem1.

**Proof:** Refer to Appendix.

**An interesting interpretation of the Hodtani region (partially improvement of the HK region ):**

In accordance with the distribution (16), we have allowed $u_1w_1$ (and also $u_2w_2$) to be correlated and used a binning scheme (see the proof of theorem 7 in Appendix). Therefore, we have added two additional terms to every rate in the HK region: One positive term $I(U_1;W_1|Q)$ or $I(U_2;W_2|Q)$ indicating the input correlation, and one negative term $-I(U_1;W_1|Q)$ or $-I(U_2;W_2|Q)$ illustrating the binning scheme. In the rates including $S_1$ or $S_2$, we have both positive and negative terms cancelling each other and there is not any difference between these rates in (3) and (17). In the rates including $T_1$ and or $T_2$, we have only the correlation resulting in additional positive term and observe a partial difference between the two regions ( compare the rates $T_1, T_2, T_1 + T_2$ in (3) and (17), specifically consider $S_i, T_i, S_i + T_i, i = 1,2$, as in Fig.3 ).

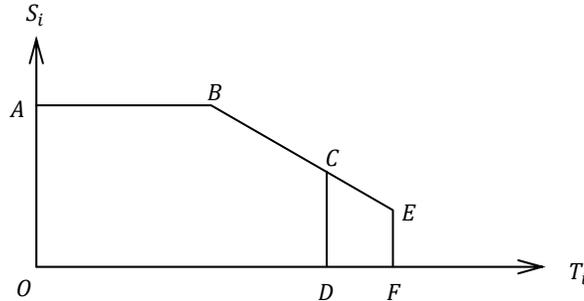

Fig.3 Shape of $(S_i, T_i)$, i=1,2 : (ABCDOA) for the HK region and (ABCEFDOA) for our region

Now the above region is described as $(R_1, R_2)$ rates.

**Theorem 8.** The $S_{IC}^{Hod}(Z)$ region in theorem 7 can be described, using the Fourier-Motzkin algorithm, as $\mathcal{R}_{Hod}$ being the set of $(R_1, R_2)$ satisfying:

$$R_1 \leq d_1 \quad (18\text{-}1)$$
$$R_1 \leq a_1 + C_2 \quad (18\text{-}2)$$
$$\boldsymbol{R_1 \leq a_1 + e_2} \quad (18\text{-}3)$$
$$R_2 \leq d_2 \quad (18\text{-}4)$$
$$R_2 \leq a_2 + C_1 \quad (18\text{-}5)$$
$$\boldsymbol{R_2 \leq a_2 + e_1} \quad (18\text{-}6)$$
$$R_1 + R_2 \leq a_2 + g_1 \quad (18\text{-}7)$$
$$R_1 + R_2 \leq a_1 + g_2 \quad (18\text{-}8)$$
$$R_1 + R_2 \leq e_1 + e_2 \quad (18\text{-}9)$$
$$2R_1 + R_2 \leq a_1 + g_1 + e_2 \quad (18\text{-}10)$$
$$\boldsymbol{2R_1 + R_2 \leq 2a_1 + e_2 + F_2} \quad (18\text{-}11)$$
$$2R_2 + R_1 \leq a_2 + g_2 + e_1 \quad (18\text{-}12)$$
$$\boldsymbol{2R_2 + R_1 \leq 2a_2 + e_1 + F_1} \quad (18\text{-}13)$$

**Proof.** This theorem is proved by virtue of the Fourier-Motzkin elimination technique the same as in theorem 2 (the details are omitted), but with two differences:

First, here, the inequalities $C_i \leq e_i$, $i = 1,2$ are not satisfied, and hence there remain $R_1 \leq a_1 + e_2$, $R_2 \leq a_2 + e_1$, in addition to the inequalities in theorem 2.

Second, in our general distribution (16), $U_i, W_i$, $i = 1,2$ are not independent given $Q$, the result of which is violating (7). Therefore, the inequalities (18-11) and (18-13) remain and are not redundant due to (18-2,7) and (18-5,8), respectively, despite the case in theorem 2 for the HK region.

**Remark 3.** Also, we can derive our $(R_1, R_2)$ region with modified error definition, analogously as in theorem 3 for the HK region. The result is 13 inequalities similar to (8-1)-(8-13) with the difference that the terms $f_i$, $i = 1,2$ in theorem 3 are replaced by $F_i$, $i = 1,2$.

**Comparing the HK and the Hodtani rate regions**

This comparison is easy. There are two differences between the two coding, that is, in the Hodtani region, first, $U_i, W_i$, $i = 1,2$ are dependent random variables given $Q$ and second, the dependence is described by random binning.

To compare the HK and the Hodtani rate regions, it is sufficient to review $S_{IC}^{HK}$ in theorem 1 and $S_{IC}^{Hod}$ in theorem 7, and also, $\mathcal{R}_{HK}$ in theorem 2 and $\mathcal{R}_{Hod}$ in theorem 8, thereby showing that the HK region is a special case of the Hodtani region:

**Lemma 3.** $S_{IC}^{HK}$ (3-1,…,14) is a special case of $S_{IC}^{Hod}$ (17-1,…,14) or
$$(T_1, S_1, T_2, S_2) \in S_{IC}^{HK} \Longrightarrow (T_1, S_1, T_2, S_2) \in S_{IC}^{Hod},$$
and the inverse is not true.

**Proof.** Due to $B_i \geq b_i$, $C_i \geq c_i$, $F_i \geq f_i$, $i = 1,2$, and the equality of other terms, the proof is obvious.

**Lemma 4.** $\mathcal{R}_{HK} = $ (4-1,…,9) or (4-1,…,13) in theorem 2 is a special case of $\mathcal{R}_{Hod} = $ (18-1,…,13) in theorem 8 or:
$$(R_1, R_2) \in \mathcal{R}_{HK} \Longrightarrow (R_1, R_2) \in \mathcal{R}_{Hod},$$
and the inverse is not true.

**Proof.** Due to $C_i \geq c_i$ and $F_i \geq f_i$, $i = 1,2$, the proof is obvious.

**Explanation of the proofs in lemmas 3 and 4:**

1. In theorems 1 and 7, $B_i \neq b_i$, $C_i \neq c_i$ and $F_i \neq f_i$, $i = 1,2$.

2. If we consider the distribution (2) instead of (16), random variables, $U_i, W_i$, $i = 1,2$ become independent given $Q$ and result in $I(U_2; W_2|Q) = I(U_1; W_1|Q) = 0$ or $B_i = b_i$, $C_i = c_i$ and $F_i = f_i$, $i = 1,2$, hence theorem 7 is reduced to theorem 1, i.e., the Hodtani region is reduced to the HK region.

3. Assuming the independence of $U_i, W_i$, $i = 1,2$ given $Q$, as in the HK region, first, we have $C_i = e_i$, $i = 1,2$, resulting in the redundancy of (18-3) and (18-6) due to (18-2) and (18-4), respectively. And second $F_i = f_i$, $i = 1,2$, as explained in the proof of theorem 2, the relations (6) and (7) are satisfied and hence (18-11), (18-13) become redundant due to (18-2,7),(18-5,8) respectively. Therefore, the thirteen relations in theorem 8 are reduced to the nine relations (4-1)-(4-9) in theorem 2, i.e., the HK region is a special case of the Hodtani region.

4. The HK region $S_{IC}^{HK}$ is an intersection of two polymatroids while in the Hodtani region, we don't have generally two polymatroids; but depending on the relations between $I(U_i; W_i|Q)$ and $I(U_i; Y_i|Q)$, $I(U_i; Y_i|QW_i)$, $i = 1,2$ we can have the polymatroids.

**Comparing the CMG and the Hodtani rate regions**

This comparison is not as easy as the comparison between the HK and Hodtani regions.

Because, first, in the Hodtani coding $U_i, W_i$, $i = 1,2$ are dependent given $Q$, and hence, for any fixed distribution (9), we don't have a corresponding distribution (16); And second, the input correlation is applied by superposition coding in the CMG region, i.e. superposition of private messages by $x_i$, $i = 1,2$ over $q$ and common messages $w_i$, $i = 1,2$, where as in the Hodtani region, the correlation is applied by binning scheme.

In other words, the two regions are not comparable directly. According to lemma 4 and the equivalencies in theorem 6, the CMG region is a special case of the Hodtani region. Specifically in accordance with the proof in lemma 2,i.e., $d_1 \geq \mathcal{D}_1$ and hence, (15-1) $\Longrightarrow$ (18-1) and the inverse is not true.

## V. CONCLUSION

By allowing the input auxiliary random variables to be correlated, we have obtained a partially improved version of the HK region for the general IC, using binning scheme and jointly decoding. Then, we have shown that the HK and hence the CMG regions are special cases of our region. In our region, every rate for the IC, has generally three terms: the first is a general HK term, the second is due to the input correlation and the third is a result of binning scheme. In another paper, we extended this generalization of input distribution to the cognitive radio channel and obtained more general theorems and results.

# APPENDIX

**The proof of theorem 7**

It is sufficient to show that any element of $S_{IC}^{Hod}(Z)$ for each $Z \in \mathcal{P}_{IC}^{Hod}$ is achievable. So, fix $Z = (QU_1W_1U_2W_2X_1X_2Y_1Y_2)$ and take any $(T_1, S_1, T_2, S_2)$ satisfying the constraints of the theorem.

**Codebook generation:** Consider $n > 0$, some distribution of the form (16) and

$p(u_1|q) = \sum_{w_1} p(w_1|q) p(u_1|qw_1)$

$p(u_2|q) = \sum_{w_2} p(w_2|q) p(u_2|qw_2)$.

Therefore, by using binning scheme we can generate sequences of $\boldsymbol{u_1}$ and $\boldsymbol{u_2}$ independently of $\boldsymbol{w_1}$ and $\boldsymbol{w_2}$. So,

1. generate a n-sequence $\boldsymbol{q}$, i.i.d. according to $\prod_{i=1}^{n} p(q_i)$, and for the codeword $\boldsymbol{q}$:
2. Generate $\lfloor 2^{nT_1} \rfloor$ conditionally independent codewords $\boldsymbol{w_1}(j)$, $j \in \{1,2,\cdots,\lfloor 2^{nT_1} \rfloor\}$ according to $\prod_{i=1}^{n} p(w_{1i}|q_i)$.
3. Generate $\lfloor 2^{ns_1} \rfloor$ ($s_1$ **is small letter**) n-sequence $\boldsymbol{u_1}(l)$, $l \in \{1,\cdots,\lfloor 2^{ns_1} \rfloor\}$, i.i.d. according to $\prod_{i=1}^{n} p(u_{1i}|q_i)$ and throw them randomly into $\lfloor 2^{nS_1} \rfloor$ ($S_1$ **is capital letter**) bins such that the sequence $\boldsymbol{u_1}(l)$ in bin $b_1$ is denoted as $\boldsymbol{u_1}(b_1, l)$, $b_1 \in \{1,\cdots,\lfloor 2^{nS_1} \rfloor\}$.
4. Generate $\lfloor 2^{nT_2} \rfloor$ n-sequence $\boldsymbol{w_2}(m)$, $m \in \{1,\cdots,\lfloor 2^{nT_2} \rfloor\}$, i.i.d. according to $\prod_{i=1}^{n} p(w_{2i}|q_i)$.
5. Generate $\lfloor 2^{ns_2} \rfloor$ ($s_2$ **is small letter**) n-sequence $\boldsymbol{u_2}(k)$, $k \in \{1,\cdots,\lfloor 2^{ns_2} \rfloor\}$, i.i.d. according to $\prod_{i=1}^{n} p(u_{2i}|q_i)$ and throw them randomly into $\lfloor 2^{nS_2} \rfloor$ ($S_2$ **is capital letter**) bins such that the sequence $\boldsymbol{u_2}(k)$ in bin $b_2$ is denoted as $\boldsymbol{u_2}(b_2, k)$, $b_2 \in \{1,\cdots,\lfloor 2^{nS_2} \rfloor\}$.

**Encoding:** The aim is to send a two dimensional message at each sender. The messages are mapped into $x_1$ and $x_2$ through deterministic encoding functions $f_1$ and $f_2$ (as in [6]). The sender $TX_1$ to send $(j, b_1)$, knowing $\boldsymbol{q}$ looks for $\boldsymbol{w_1}(j)$ and finds a sequence $\boldsymbol{u_1}(b_1, l)$ in bin $b_1$ such that $(\boldsymbol{q}, \boldsymbol{w_1}(j), \boldsymbol{u_1}(b_1, l)) \in A_\varepsilon^n$; then generates $x_1$ i.i.d. according to $x_{1i} = f_1(w_{1i}(j), u_{1i}(b_1, l)|q_i)$, $i = 1,\cdots, n$; and sends it. The sender $TX_2$ to send $(m, b_2)$, knowing $\boldsymbol{q}$ looks for $\boldsymbol{w_2}(m)$ and finds a sequence $\boldsymbol{u_2}(k)$ in bin $b_2$ such that $(\boldsymbol{q}, \boldsymbol{w_2}(m), \boldsymbol{u_2}(b_2, k)) \in A_\varepsilon^n$; then generates $x_2$ i.i.d. according to $x_{2i} = f_2(w_{2i}(m), u_{2i}(b_2, k)|q_i)$, $i = 1,\cdots, n$ and sends it.

**Decoding and analysis of error probability:** The receivers $RX_1$ and $RX_2$ decode the corresponding messages, based on strong joint typicality [6]. It is assumed that all messages are equiprobable. Without loss of generality, we may confine ourselves to the situation where $(j = 1, b_1 = 1; m = 1, b_2 = 1)$ was sent.

The receiver $RX_1$, by receiving $\boldsymbol{y_1}$ and knowing $\boldsymbol{q}$, decodes $j = 1, b_1 = 1, m = 1$ or $j(b_1, l)\, m = 1(1, l)1$ simultaneously [6]. We define the event $E_{j(b_1,l)\,m}$ and $P_{e_1}^n$ as follows.

$E_{j(b_1,l)m} = \{(\boldsymbol{q}, \boldsymbol{w_1}(j), \boldsymbol{u_1}(b_1, l), \boldsymbol{w_2}(m), \boldsymbol{y_1}) \in A_\varepsilon^n\}$

$P_{e_1}^{(n)} = P\{E_{1(1,l)1}^C \cup E_{j(b_1,l)m \neq 1(1,l)1}\} \leq P(E_{1(1,l)1}^C) + \sum_{j(b_1,l)m \neq 1(1,l)1} P(E_{j(b_1,l)m}) \leq \varepsilon + \sum_{j \neq 1, b_1 = m = 1}^{\boxed{1}} \cdots +$

$\sum_{b_1 \neq 1, j = m = 1}^{\boxed{2}} \cdots + \sum_{m \neq 1, j = b_1 = 1}^{\boxed{3}} \cdots + \sum_{j \neq 1, m \neq 1, b_1 = 1}^{\boxed{4}} \cdots + \sum_{j \neq 1, b_1 \neq 1, m = 1}^{\boxed{5}} \cdots + \sum_{m \neq 1, b_1 \neq 1, j = 1}^{\boxed{6}} \cdots + \sum_{j \neq 1, m \neq 1, b_1 \neq 1}^{\boxed{7}} \cdots$

Let us choose $\boxed{1}, \boxed{2}, \boxed{7}$ for evaluation; in accordance with the codebook generation and the original distribution (16) we have:

- $\sum_{j \neq 1, b_1 = m = 1}^{\boxed{1}} \cdots \leq 2^{nT_1} (p(\boldsymbol{q}, \boldsymbol{w_1}(j), \boldsymbol{u_1}(1, l), \boldsymbol{w_2}(1), \boldsymbol{y_1}) \in A_\varepsilon^n) \leq$

  $2^{nT_1} \sum_{(\boldsymbol{q}, \boldsymbol{w_1}(j), \boldsymbol{u_1}(1,l), \boldsymbol{w_2}(1), \boldsymbol{y_1}) \in A_\varepsilon^n} p(\boldsymbol{q}, \boldsymbol{w_1}(j), \boldsymbol{u_1}(1, l), \boldsymbol{w_2}(1), \boldsymbol{y_1}) \leq$

  $2^{nT_1} \|A_\varepsilon^n\| p(\boldsymbol{q})p(\boldsymbol{w_1}|\boldsymbol{q})p(\boldsymbol{u_1}|\boldsymbol{q})p(\boldsymbol{w_2}|\boldsymbol{q})p(\boldsymbol{y_1}|\boldsymbol{q}u_1w_2) \leq 2^{nT_1} \cdot 2^{nH(QW_1U_1W_2Y_1)} \cdot 2^{-nH(Q)} \cdot 2^{-nH(W_1|Q)} \cdot$

  $2^{-nH(U_1|Q)} \cdot 2^{-nH(W_2|Q)} \cdot 2^{-nH(Y_1|QU_1W_2)} = 2^{-n(I(U_1;W_1|Q) + I(Y_1;W_1|QW_2U_1) - T_1)}$

- $\sum_{b_1 \neq 1, j = m = 1}^{\boxed{2}} \cdots \leq 2^{ns_1} (p(\boldsymbol{q}, \boldsymbol{w_1}(1), \boldsymbol{u_1}(b_1, l), \boldsymbol{w_2}(1), \boldsymbol{y_1}) \in A_\varepsilon^n) \leq$

  $2^{ns_1} \|A_\varepsilon^n\| p(\boldsymbol{q})p(\boldsymbol{w_1}|\boldsymbol{q})p(\boldsymbol{u_1}|\boldsymbol{q})p(\boldsymbol{w_2}|\boldsymbol{q})p(\boldsymbol{y_1}|\boldsymbol{q}w_1w_2) \leq \cdots = 2^{-n(I(U_1;W_1|Q) + I(Y_1;U_1|QW_1W_2) - s_1)}$

- $\sum_{j \neq 1, m \neq 1, b_1 \neq 1}^{\boxed{7}} \cdots \leq 2^{n(s_1 + T_1 + T_2)} (p(\boldsymbol{q}, \boldsymbol{w_1}(j), \boldsymbol{u_1}(b_1, l), \boldsymbol{w_2}(m), \boldsymbol{y_1}) \in A_\varepsilon^n) \leq$

  $2^{n(s_1 + T_1 + T_2)} \|A_\varepsilon^n\| p(\boldsymbol{q})p(\boldsymbol{w_1}|\boldsymbol{q})p(\boldsymbol{u_1}|\boldsymbol{q})p(\boldsymbol{w_2}|\boldsymbol{q})p(\boldsymbol{y_1}|\boldsymbol{q}) \leq \cdots = 2^{-n(I(U_1;W_1|Q) + I(Y_1;W_1W_2U_1|Q) - s_1 - T_1 - T_2)}$.

Similarly, the other terms can be evaluated. In order to ($P_{e_1}^{(n)} \to 0$ as the block length $n \to \infty$), it is necessary and sufficient that:

$$\begin{cases} s_1 \le I(U_1;W_1|Q) + I(Y_1;U_1|QW_1W_2) \\ T_1 \le I(Y_1;W_1|W_2U_1Q) + I(U_1;W_1|Q) \\ T_2 \le I(U_1;W_1|Q) + I(Y_1;W_2|QW_1U_1) \\ s_1 + T_1 \le I(Y_1;U_1W_1|QW_2) + I(U_1;W_1|Q) \\ s_1 + T_2 \le I(Y_1;U_1W_2|QW_1) + I(U_1;W_1|Q) \\ T_1 + T_2 \le I(Y_1;W_1W_2|QU_1) + I(U_1;W_1|Q) \\ s_1 + T_1 + T_2 \le I(Y_1;U_1W_1W_2|Q) + I(U_1;W_1|Q) \end{cases} \quad (A_1), (s_1 \text{ is small letter})$$

from where, considering the binning condition:
$I(U_1;W_1|Q) \le s_1 - S_1$ or $S_1 - s_1 \le -I(U_1;W_1|Q)$,
the relations ($A_1$) yield to the constraints (17-1)-(17-7) in theorem 7.

Error probability analysis for the receiver $RX_2$ can be done similarly and the inequalities (17-8)-(17-14) can be proved (for brevity, the details are omitted).